\documentclass[reprint,
superscriptaddress,
amsmath,
amssymb,
aps,
pra,
longbibliography]{revtex4-2}

\usepackage[T1]{fontenc}
\usepackage[utf8]{inputenc}

\usepackage{times}
\usepackage[unicode=true,breaklinks=true,colorlinks=true]{hyperref}
\hypersetup{citecolor=blue,urlcolor=blue}
\usepackage{graphicx}
\usepackage{bm}
\usepackage{bbold}
\usepackage{ulem}

\usepackage{mathtools}
\usepackage{tensor}

\newcommand{\rev}[1]{\textcolor{black}{#1}}

\begin{document}

\title{Second-order topological insulator in periodically driven lattice}

\author{Ying Lei}
\affiliation{School of Physics, Huazhong University of Science and Technology, Wuhan 430074, China}
\author{Xi-Wang Luo}
\affiliation{Key Laboratory of Quantum Information, University of Science and Technology of China, Hefei, Anhui 230026, China}
\author{Shaoliang Zhang}
\email{shaoliang@hust.edu.cn}
\affiliation{School of Physics, Huazhong University of Science and Technology, Wuhan 430074, China}
\affiliation{State Key Laboratory of Quantum Optics and Quantum Optics Devices, Shanxi University, Taiyuan 030006,China}

\date{\today}

\begin{abstract}

The higher-order topological insulator (HOTI) is a new type of topological system which has special bulk-edge correspondence compared with conventional topological insulators.
In this work, we propose a scheme to realize Floquet HOTI in ultracold atom systems. With the combination of periodically spin-dependent  driving of the superlattices and a next-next-nearest-neighbor $d$-wave-like anisotropic coupling term between different spin components, a Floquet second-order topological insulator with four zero-energy corner states emerges, {\color{black}whose Wannier bands are gapless and exhibit interesting bulk topology. Furthermore,} the anisotropic coupling with nearest-neighbor form will also induce some \rev{intriguing} topological phenomena, e.g. non-topologically protected corner states and topological semimetal for two different types of lattice structures respectively. Our scheme may give insight into the construction of different types of higher-order topological insulators in synthetic systems. It also provides an experimentally feasible platform to research the relations between different types of topological states and may have a wide range of applications in future.
\end{abstract}

\maketitle

\section{Introduction}

In the last few years, higher-order topological insulator, as a new type of topological matters with unconventional bulk-edge correspondence, has attracted intensive interests \cite{Bernevig1,Bernevig2,Schindler,HOTI1,HOTI3,HOTI4,HOTI5,HOTI6,HOTI7,HOTI8,HOTI9,HOTI10,HOTI11,HOTI12,HOTI13,HOTI14} in communities of topological physics and condensed matter physics. For example, $n$-dimensional second-order topological insulator (SOTI) has topological protected gapless $(n-2)$-dimensional edge states while the $(n-1)$-dimensional boundaries are gapped, which is very different from conventional topological matters with topological protected gapless $(n-1)$-dimensional boundaries. People have proposed considerable interesting models to construct HOTI. One of the most prominent proposals is the extension of one-dimensional (1D) Su-Schrieffer-Heeger (SSH) model \cite{SSHmodel} to \rev{the} two-dimensional (2D) system \cite{Bernevig1,Bernevig2}. With \rev{an} additional $\pi$ flux threaded through each plaquette, the SOTI can be constructed and the topological properties are characterized by the quantized quadrupole moment and edge polarization in the system. Another \rev{important} proposal is based on a 2D topological insulator (TI)  such as Bernervig-Hughes-Zhang (BHZ) model with an additional $d$-wave-like anisotropic coupling term \cite{ZhongboYan}. In \rev{this models}, the emergence of topologically protected zero-energy corner states can be explained by the low-energy effective theory. In addition, the concept of HOTI is extended into many-body systems, and higher-order topological superconductors are proposed \cite{superconductors0,superconductors1,superconductors2,superconductors3,superconductors4}, which support Majorana zero modes\cite{majorana1,majorana2,majorana3} that can be potentially applied to topological quantum computation.

Recently, the research about HOTI is developing very rapidly. A lot of schemes to construct HOTI is put forward in different types of materials in condensed matter physics, \rev{and relevant experimental phenomena have been observed, e.g., in bismuth \cite{bismuth}. On the other hand, zero-energy corner states have \rev{also} been realized experimentally \cite{photonic,electriccircuit,microwave} in many synthetic systems such as photonic crystals, electronic circuits and microwave resonators.} \rev{Remarkably, ultracold atom system is proposed to be an ideal synthetic platform to simulate many fascinating topological phenomena}. Due to its highly controllable properties, not only those theoretically proposed topological models can be realized in experiments, but also some new types of topological matters and other intriguing quantum phenomena might be predicted and explored in ultracold atom system. \rev{In particular,} {\color{black} direct measurements of bulk topology are accessible for most experiments using ultracold atom  \cite{tomo1,tomo2}}, \rev{as compared with other synthetic systems.}

In this work, we propose an experimentally feasible model of SOTI based on ultracold atoms in optical lattices. We confine ultracold atoms which \rev{have} two hyperfine spin states in a checkerboard lattice. By applying  a spin-dependent circular driving \cite{shaking1,shaking3}, the whole system can \rev{be regarded} as an effective \rev{2D TI} \cite{QSHE1,QSHE2,QSHE3} with gapless edge states which are protected by Z2 symmetry. Then by adding an additional next-next-nearest-neighbor (NNNN) anisotropic coupling between different spin components, the edge states will be gapped out and topologically protected zero-energy corner states arise at the intersection of adjacent boundaries. {\color{black} Different from earlier works, the Wannier bands for characterizing the bulk topology are gapless. \rev{However}, the topological invariant in the bulk can still be well-defined \rev{due to the reflection symmetries} in our discussion.} {\color{black}In this work, the coupling between different spin components has \rev{a} $d$-wave form instead of the chiral $d+id$ \rev{one}, \rev{thus} the next-nearest-neighbor (NNN) coupling is irrelevant and don't need to be considered.} Furthermore, when the NNNN anisotropic coupling term is replaced by the conventional nearest-neighbor (NN) anisotropic coupling, one can also find that there exist interesting topological phenomena, such as non-topologically protected corner states and topological semimetal \cite{semimetal1,semimetal2,semimetal3} for different NN coupling terms respectively. This work provides a new perspective to understand the relationship between the choice of anisotropic coupling term and the emergence of higher-order topological insulator. \rev{In addition,} the proposed model enriches theoretical research about HOTI and may have a wide range of applications in future.

This paper is organized as follows. In Section II, we introduce our model and its topological properties. We also give an explanation about the emergence of zero-energy corner states with \rev{an} effective edge theory and the bulk-edge-corner correspondence. In Section III, topological phenomena induced by the two types of NN coupling term are discussed, i.e. non-topologically protected corner states and topological semimetal with one or two pairs of nodal points. In Section IV, we extend our model into \rev{a} three-dimensional system and discuss its topological properties. In \rev{such a} system, chiral hinge modes and the swing of corner states can be observed. Section V is devoted to {\color{black}the proposal of experimental setup and measurements. Section VI is a conclusion}.

\section{The model and topological properties}

\begin{figure}[tph]
\centering
\includegraphics[width=83mm]{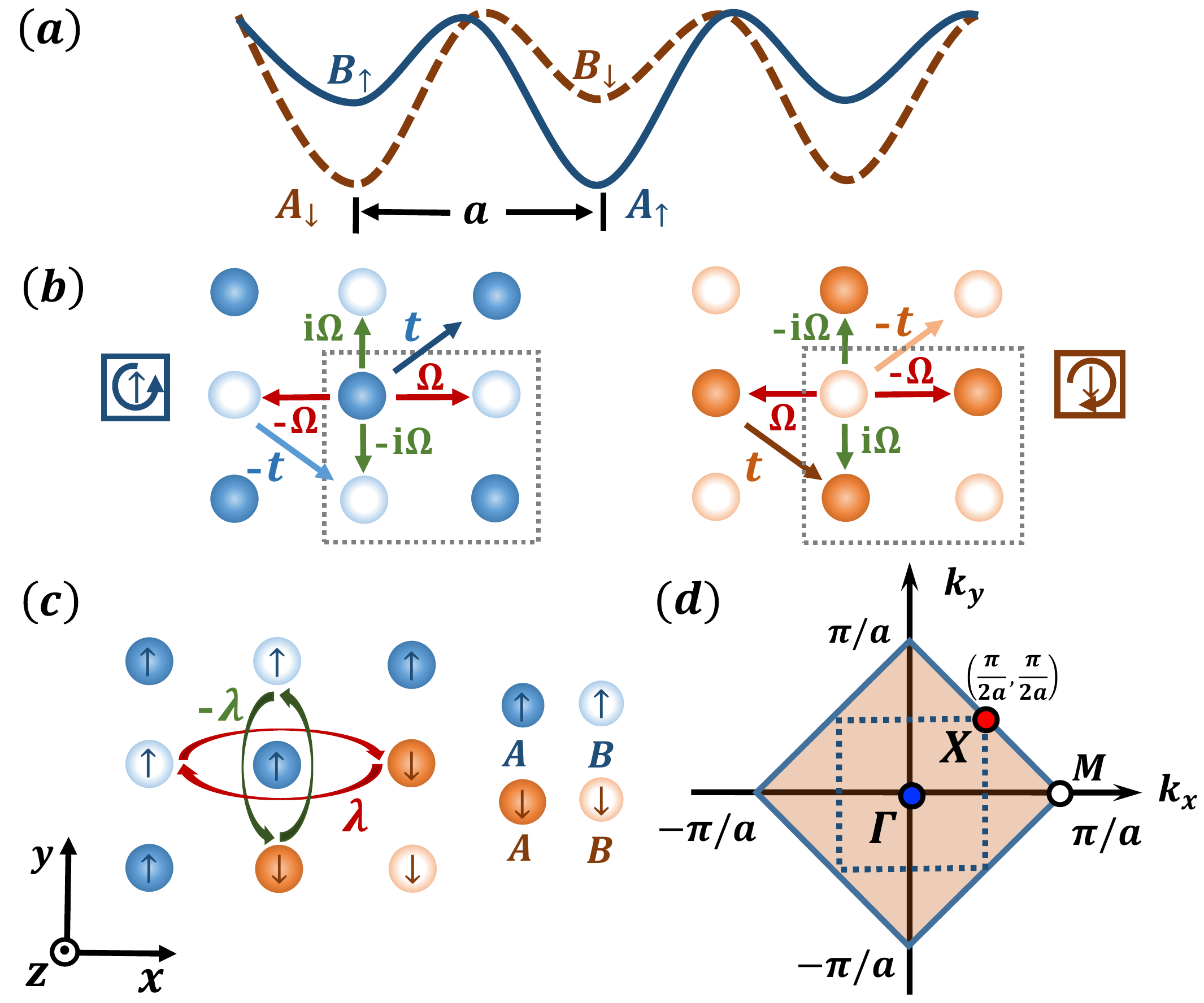}
\caption{Scheme of 2D TI model with open boundary along \rev{the} ABA direction. (a) The lattice potentials \rev{for different spin components} along \rev{the} $x$ direction, where $A$ and $B$ \rev{represent} sublattices with \rev{an} energy detuning. Blue solid lines and orange dotted lines \rev{are associated with} lattice structures \rev{for} spin-up and spin-down \rev{components} respectively. (b) The lattice structure. Shaking directions are spin-dependent, i.e. the directions for spin-up and spin down are anti-clockwise and clockwise respectively. Dark and light colors represent $A$ and $B$ sublattices respectively. The nearest-neighbor tunnelings have opposite chirality because of the spin-dependent shaking.  (c) Anisotropic coupling term. NNNN coupling between different spin components along \rev{the $x$ and $y$} directions are $\lambda$ and $-\lambda$ respectively. (d) The colored area is the Brillouin zone (BZ) of the 2D TI model. The square surrounded by the dash line is the folded BZ. }\label{model}
\end{figure}

The model we discuss in this work is based on a spin-dependent two-dimensional  checkerboard lattice, the spin-up and spin-down components of which are shifted from each other by length $a$ along the $x$ or $y$ direction, where $a$ is the displacement from site $A$ to its nearest-neighbor (NN) site $B$, as depicted in Fig.\ref{model}(a) and (b). Due to the energy detuning between \rev{the} sublattices A and B, the tunnelings between nearest-neighbor(NN) sites are forbidden. Then by applying a synthetic gauge fields \cite{slzhang} with opposite directions of circular shaking for different spin components, one can induce an unconventional NN tunneling with opposite chirality for different spin components and an effective next-nearest-neighbor (NNN) tunneling with opposite signs for A and B sublattices respectively, as shown in Fig.\ref{model}. Additionally, by adding an NNNN anisotropic coupling between different spin components, see in Fig.\ref{model}(c), the zero-energy corner states \rev{appear}, which signifies the emergence of SOTI.  According to the Floquet theory in Ref. \cite{slzhang}, the tight-binding Hamiltonian of our system in \rev{the} momentum space \rev{can be described as} $H=\sum_{\bf k}\Psi^\dag_{\bf k}H_{\bf k}\Psi_{\bf k}$ with $\Psi^\dag_{\bf k}=\big[a^\dag_{{\bf k},\uparrow},b^\dag_{{\bf k},\uparrow},a^\dag_{{\bf k},\downarrow},b^\dag_{{\bf k},\downarrow}\big]^\mathrm{T}$ and
\begin{equation}
\begin{split}
H_{\bf k}=&\big\{\delta+4t\cos(k_x a)\cos(k_ya)\big\}\sigma_z+2\Omega\sin(k_x a)\sigma_y\\
+&2\Omega\sin(k_ya)\sigma_x s_z+2\lambda\big\{\cos(2k_x a)-\cos(2k_ya)\big\}\sigma_x s_x
\label{effhamsocmanybody}
\end{split}
\end{equation}
where  $s_i, \sigma_i$ ($i=x,y,z$) denote the Pauli matrices acting on \rev{the} hyperfine spin ($\uparrow$, $\downarrow$) and orbit ($A$, $B$) degree of freedoms respectively, $2\delta$ is the effective energy detuning between sublattices $A$ and $B$, $\Omega$ and $t$ are the amplitude of NN and NNN tunneling respectively, and $\lambda$ is the amplitude of the NNNN coupling. In this work, we set $t>0$ for convention.

\begin{figure}[tph]
\centering
\includegraphics[width=85mm]{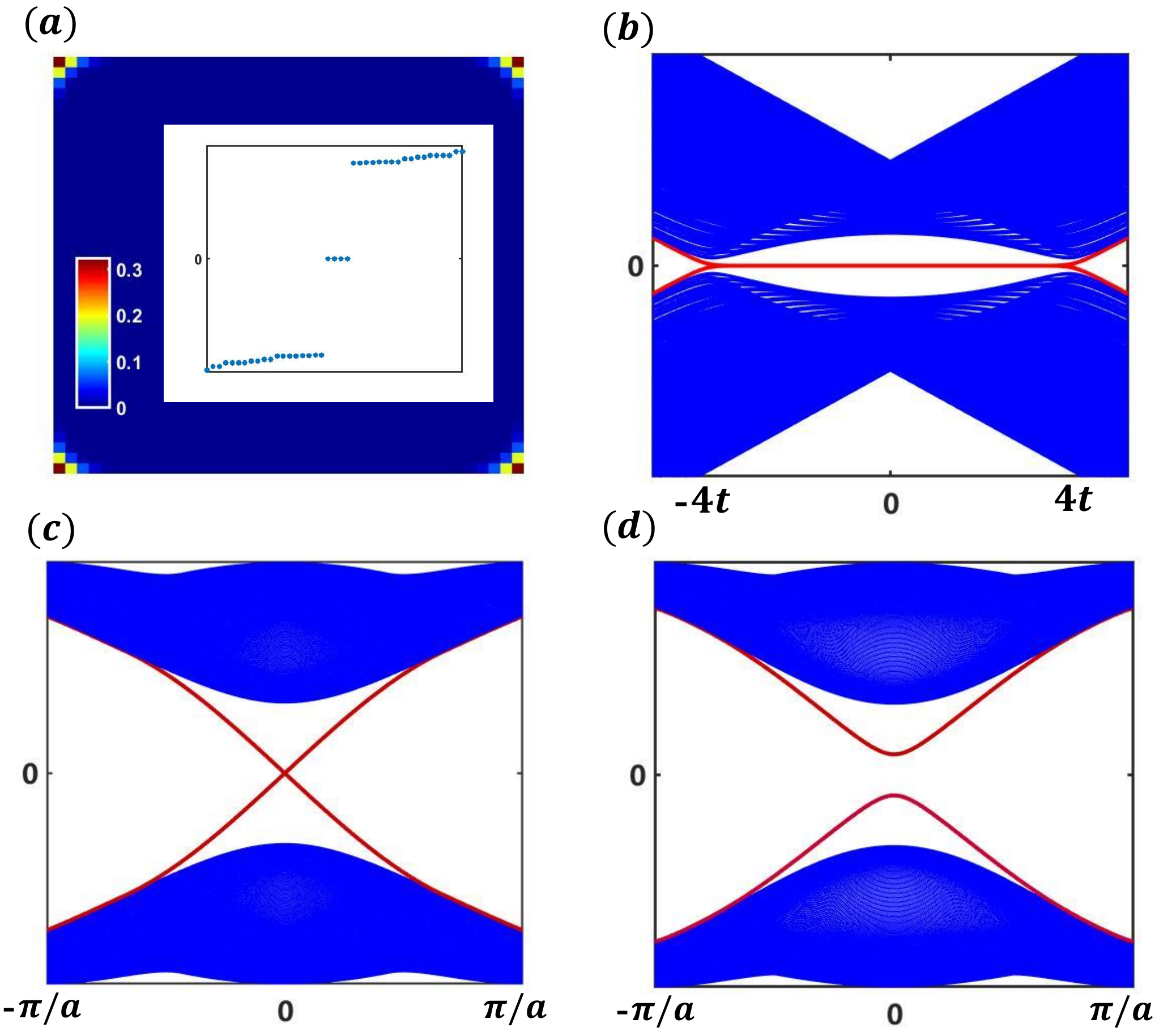}
\caption{(a) Probability density distributions of zero-energy corner states with \rev{a} lattice size $L_x\times L_y=40\times40$. The inset shows energy spectrum of the whole system. The parameter of this \rev{panel} is \rev{set as} $\delta=t$, $\Omega=t$ and $\lambda=t$. (b) Energy spectrum with  \rev{values of} effective detuning $\delta$.  (c) Energy-band structure of the system with \rev{an} open boundary along \rev{the} $x$ direction when $\lambda=0$.  (d) Energy-band structure of the system with \rev{an} open boundary along \rev{the} $x$ direction when $\lambda\ne 0$. The red curves denote two degenerate edge states.}\label{fig:cornerstate}
\label{corner}
\end{figure}

At $\lambda=0$, the Hamiltonian in Eq.(\ref{effhamsocmanybody}) and  its topological properties are very similar  to the BHZ model. We remark that in contrast to the system with a magnetic field breaking the time-reversal symmetry, there exists quantum spin Hall effect (QSHE) in  the parameter space of $\lvert \delta \rvert<4t$, i.e. a topological insulator with two opposite quantum Hall phases, the Chern numbers of which satisfy $C_{\uparrow}=-C_{\downarrow}$. According to the band structure with open boundary along the $x$ (or $y$) direction as shown in Fig.\ref{corner}(c), there exists two pairs of edge states that propagate with opposite directions. The Hamiltonian  (\ref{effhamsocmanybody}) preserves time reversal symmetry and inversion symmetry with corresponding symmetry operators $\mathcal{T}=is_y\mathcal{K}$  ($\mathcal{K}$ denoting the operation of complex conjugate) and $\mathcal{I}=\sigma_z$ respectively. So we have
$\mathcal{T}H_{\bf k}\mathcal{T}^{-1}=\mathcal{H}_{-{\bf k}}$,   $\mathcal{I}H_{\bf k}\mathcal{I}^{-1}=H_{-{\bf k}}$.


For $\lambda \neq 0$, the anisotropic coupling term breaks both the inversion symmetry and time-reversal symmetry but respect their combination $\mathcal{I}\mathcal{T}$, that is $(\mathcal{I}\mathcal{T})H_{\bf k}(\mathcal{I}\mathcal{T})^{-1}=H_{\bf k}$. One can easily verify that $(\mathcal{I}\mathcal{T})^2|\psi\rangle=-|\psi\rangle$ is satisfied for arbitrary states $|\psi\rangle$, which means that the system has a general Kramers degeneracy and energy bands in whole Brillouin zone (BZ) have \rev{a} two-fold degeneracy. By exact numerical calculation with open boundary conditions along the $x$ and $y$ direction, we find that four zero-energy states emerge and are located at the corners of 2D lattice, see Fig.\ref{corner}(a). The four zero-energy corner states are topologically protected when $\lvert \delta \rvert<4t$, as shown in Fig.\ref{corner}(b). The anisotropic coupling term gap out the energy bands on edge, as shown in Fig.\ref{corner}(d).

The emergence of zero-energy corner states can be revealed from the low-energy effective theory \cite{ZhongboYan}. In this system, the first BZ is a square. At the topological phase boundary, the band gap closes at $\Gamma=(0,0)$ and $M=(\pi/a,0)$ when $\delta=-4t$ and $\delta=4t$ respectively, as shown in Fig.\ref{model}(c). When $\delta\rightarrow -4t$, the minimum of \rev{the} gap in the whole BZ is located at $\Gamma$ point. Then the Hamiltonian (\ref{effhamsocmanybody}) can be expanded around $\Gamma$ point with the low-energy form as
\begin{equation}
\begin{split}
H_\Gamma({\bf k})=&\big\{m-2t(k^2_x a^2+k^2_ya^2)\big\}\sigma_z+2\Omega k_x a\sigma_y\\
+&2\Omega k_ya\sigma_x s_z+8\lambda(k^2_ya^2-k^2_x a^2)\sigma_x s_x
\label{approachformgamma}
\end{split}
\end{equation}
where $m=\delta+4t$. We choose the open boundary condition $y>0$, under which $k_y$ is replaced by $-i\partial_y$, and ignore the higher order terms of $k_x$,  the effective Hamiltonian can be rewritten as $H_\Gamma(k_x,-i\partial_y)=H_0+H_p$ with
 \begin{equation}
H_0=\big\{m+2ta^2\partial^2_y)\big\}\sigma_z
-2i\Omega a\partial_y\sigma_x s_z
\label{edgeham0}
\end{equation}
and \rev{a perturbation term}
 \begin{equation}
H_p=2\Omega ak_x\sigma_y-4\lambda a^2\partial^2_y\sigma_x s_x
\end{equation}
\rev{Further,} $H_0$ can be block diagonalized and the two zero-energy solutions are \rev{given by}
\begin{equation}
\psi_\uparrow\propto A(y)\left(\begin{array}{c} 1 \\ 0 \\ i \\ 0 \end{array}\right),\psi_\downarrow\propto A(y)\left(\begin{array}{c} 0 \\ 1 \\ 0 \\ -i \end{array}\right)
\label{zeroenergystate0}
\end{equation}
where $A(y)=e^{-C_+y}-e^{-C_-y}$ and $C_\pm=\frac{\Omega}{2ta}\pm\frac{\Omega}{2ta}\sqrt{1-\frac{2mt}{\Omega^2}}$. So when $\delta>-4t$, $m>0$, both $C_+$ and $C_-$ are positive, which means \rev{that} the zero-energy modes are localized on the edge. When $\delta<-4t$, we have $m<0$ and $C_-<0$, the solution is divergent, \rev{implying} that there is no zero-energy edge modes and the system is topologically trivial.  \rev{Then,} we project $H_p$ into the subspace of these zero-energy edge modes with the form
\begin{equation}
H_{y+}(k_x)\propto D_{y+}k_x s_z-M_{y+} s_y
\label{edgehamy}
\end{equation}
 where $D_{y+}=4\Omega a\int^\infty_0A^2(y)d y$ and $M_{y+}=8\lambda a^2\int^\infty_0A(y)\partial^2_y A(y)d y$. In the effective edge Hamiltonian (\ref{edgehamy}), the last term $M_{y+}$ can be viewed as an effective mass. One can use the same method to expand the effective Hamiltonian (\ref{approachformgamma}) on the other three edges as \rev{demonstrated} in Ref. \cite{ZhongboYan} and obtain \rev{three other} edge Hamiltonians of \rev{a} similar form. The effective masses on adjacent edges always have opposite signs, which means that the zero-energy "domain wall" emerges on the junction of adjacent edges according to the Jackiw-Rebbi theory \cite{Jackiw-Rebbi} when the boundary of system is a square cross section.

{\color{black} In conventional topological systems, the bulk-edge correspondence is important for verifying their topological properties. The emergence of gapless edge states always corresponds to \rev{a} nontrivial bulk topology. \rev{For HOTI}, bulk-edge-corner correspondence \rev{also exists, i.e.} the emergence of zero-energy corner states are associated with edge polarization and non-zero quantized quadrupole moment in the bulk \cite{Bernevig1,Bernevig2}.} \rev{But} in our system, because the open boundary \rev{that} we choose is along the diagonal lines of \rev{the} BZ, the choice of the above BZ may not be suitable to explore the topological properties in the bulk. This can also be seen by noticing the zero-energy edge mode solution Eq.\ref{zeroenergystate0}, which is located at the edge \rev{of} $y=0$, and thus
the lattice period in this subspace is $2a$ along \rev{the} $x$ direction, reducing the Bloch momentum $k_x$ within the dashed square in Fig.\ref{model}(d). To overcome this difficulty, we can enlarge the size of unit cell to contain four nearest-neighbor sites, as shown in Fig.\ref{model}(b). Then the BZ will be folded into the range \rev{enclosed} by the dash line in Fig. \ref{model}(d) and the corresponding Hamiltonian in the momentum space can be rewritten as
\begin{equation}
H_{\bf k}=H^0_{\bf k}s_z+2\lambda\big\{\cos(k_x d)-\cos(k_yd)\big\}s_x
\label{hamfolded}
\end{equation}
where \rev{the Hamiltonian $H^0_{\bf k}$} of spin-up component can be written as
\begin{equation}\label{ham_spin_up}
\begin{split}
 H^0_{\bf k}&=\delta\sigma_z\tau_z+\Omega\big\{1-\cos(k_x d)\big\}\sigma_z\tau_x-\Omega\sin(k_x d)\sigma_z\tau_y \\
 -&\Omega\sin(k_yd)\sigma_x-\Omega\big\{1-\cos(k_yd)\big\}\sigma_y \\
-&t\sin(k_x d)\sin(k_yd)\sigma_x\tau_x \\
-&t\big\{1+\cos(k_x d)+\cos(k_yd)+\cos(k_x d)\cos(k_yd)\big\}\sigma_y\tau_y \\
+&t\big\{\sin(k_x d)+\sin(k_x d)\cos(k_yd)\big\}\sigma_y\tau_x \\
+&t\big\{\sin(k_yd)+\sin(k_yd)\cos(k_x d)\big\}\sigma_x\tau_y
\end{split}
\end{equation}
Here $d=2a$ is the length of the enlarged unit cell. $s_i$ is still Pauli matrices of hyperfine spin states ($\uparrow$, $\downarrow$). Both $\sigma_i$ and $\tau_i$ correspond to Pauli matrices of orbit degree of freedom. Then the open boundary can be chosen along the edges of the folded BZ. {\color{black} The Hamiltonian (\ref{hamfolded}) in the folded BZ respects reflection symmetries along the $x$ and $y$ directions with corresponding symmetry operators $M_x=\sigma_z\tau_y s_x$ and $M_y=\sigma_x s_x$ respectively, which \rev{anticommute with each other, i.e. $\{M_x,M_y\}=0$}. In general, the quantization of edge polarization and quadrupole moment are both protected by the reflection symmetries.}

\begin{figure}[tph]
\centering
\includegraphics[width=84mm]{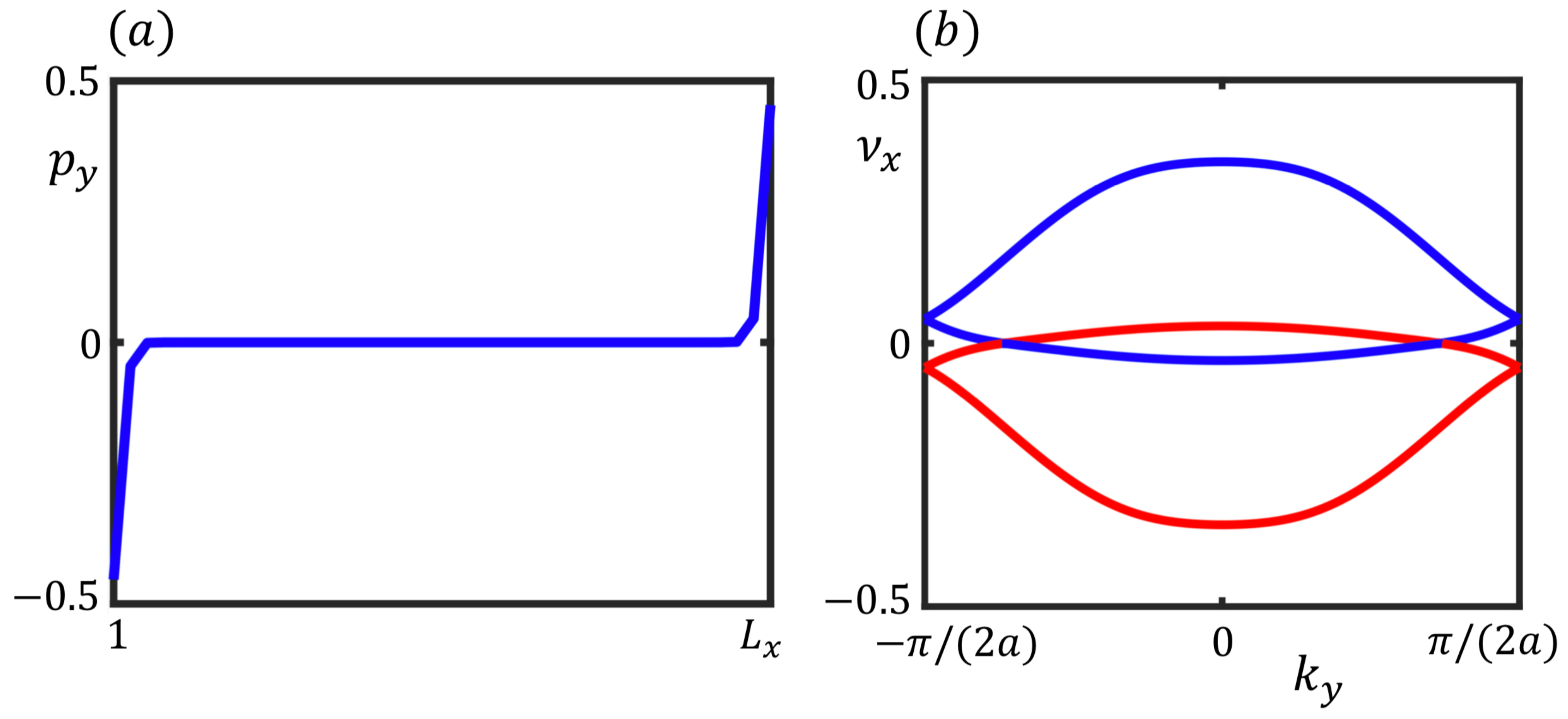}
\caption{ (a) Edge polarization $p_y$ \rev{with a} lattice size along \rev{the} $x$ direction \rev{as} $L_x=40$. (b) Wannier bands for different \rev{values} of $k_y$. {\color{black} Red and blue lines correspond to two different types of Wannier bands with eigenvalues $\nu^+_x$ and $\nu^-_x$ respectively. } The parameters are chosen as $\delta=0.4t$, $\omega=0.8t$ and $\lambda=0.3t$.}\label{fig:edgebulk}
\end{figure}

{\color{black}
One can choose \rev{an} open boundary along the $x$ or $y$ direction and estimate the edge polarization $p^\mathrm{edge}_x$ and $p^\mathrm{edge}_y$ respectively \cite{Bernevig1,Bernevig2}. As in Fig.\,\ref{fig:edgebulk}(a), the edge polarization is quantized and non-zero when $|\delta|<4t$, and a sudden change to zero can also be observed when $|\delta|>4t$, which means \rev{that} our system has an edge-corner correspondence.

The bulk topology is interesting in our system. The Wannier bands $\nu_x$ as the set of Wannier centers along $x$ as a function of $k_y$ can be estimated by calculating the Wilson loop along the $x$ direction in the folded BZ \cite{Bernevig1,Bernevig2}. As shown in Fig.\,\ref{fig:edgebulk}(b), these Wannier bands are gapless, which is very different from \rev{the scenarios} in Ref.\cite{Bernevig1,Bernevig2}. But because of \rev{the presence of} the reflection symmetry, the distribution of these Wannier bands still satisfy the relation $\nu^+_x=-\nu^-_x$. One can choose \rev{one} Wannier sector, for example $\nu^+_x$, to estimate the quadrupole moment by calculating the nested Wilson loop, we just need to integrate along the Wannier sector $\nu_x^+$. We can calculate the non-Abelian nested Wilson loop of the two Wannier bands, instead of integrate them in turns. The result shows that the quadrupole moment is quantized and its value is $1/2$ when $|\delta|<4t$, which \rev{indicates that} our system has a bulk-edge-corner correspondence. Our results also \rev{demonstrate} that even \rev{when} the Wannier bands is gapless, the quantized quadrupole moment can still be well-defined \rev{if} the system has corresponding symmetries.}

It's worth pointing out that the correspondence between the quantized edge polarization and zero corner modes is an intrinsic property of the system and it is irrelevant with the choice of the BZ. This can be verified by adding a very small perturbation to distinguish the amplitude of intra-cell and inter-cell NN tunneling $\Omega$, which would make the enlarged unit cell a primitive cell. We find that the topological properties \rev{is unchanged}, which means that the above topological characterization is feasible and valid.

\section{Topological properties of system with nearest-neighbor anisotropic coupling}

\begin{figure}[tph]
\centering
\includegraphics[width=88mm]{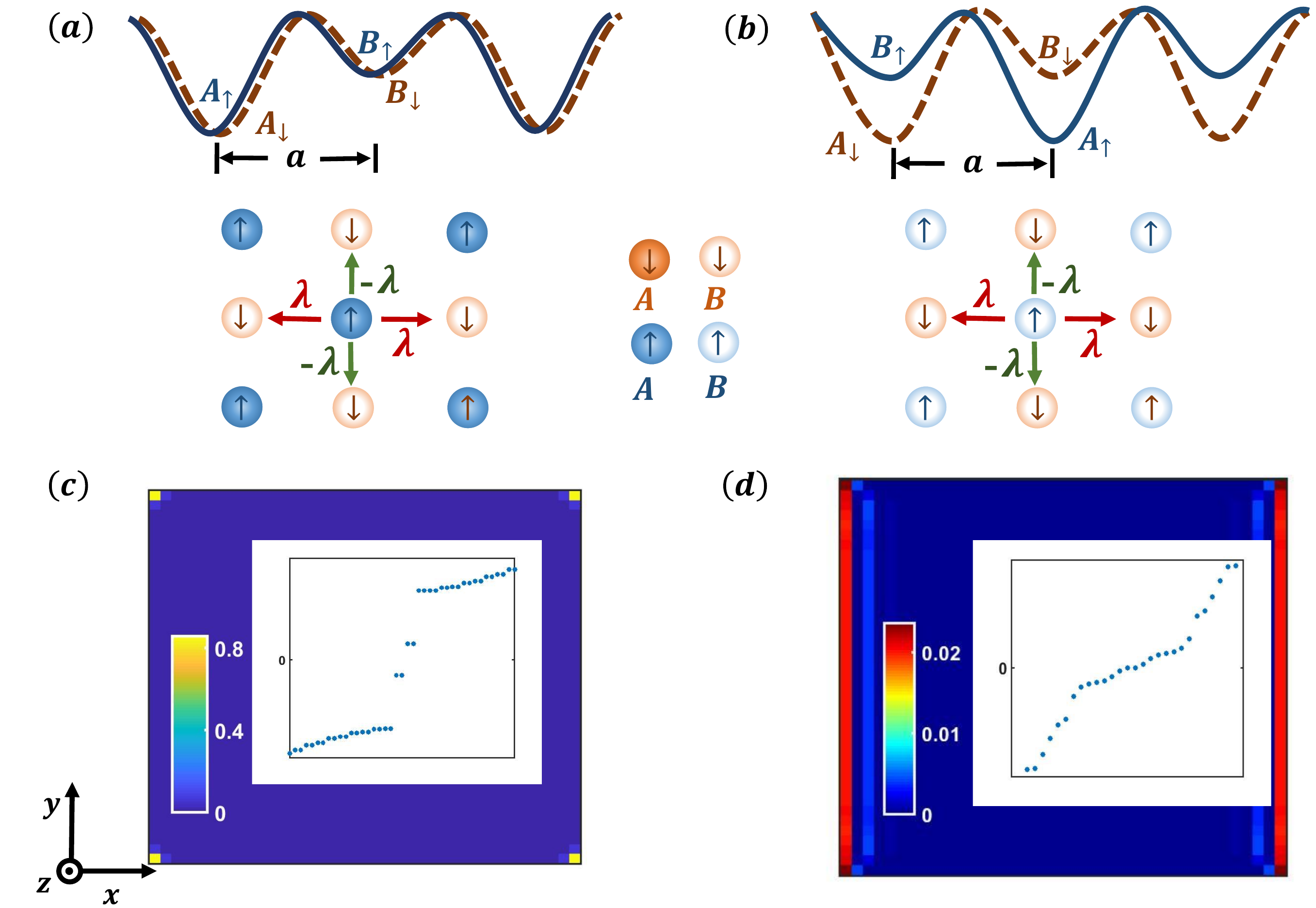}
\caption{\rev{The upper panels show the spin-independent lattice structures (a) for Hamiltonian (\ref{effhamwithoutglide}) and (b) for Hamiltonian (\ref{effhamwithouttimereversal}) as well as their corresponding NN coupling terms. The lower panels (c) and (d) depict the density distributions of nonzero-energy corner states in the system with Hamiltonians (\ref{effhamwithoutglide}) and  (\ref{effhamwithouttimereversal}) respectively. The inset shows the energy spectrum of whole system. The parameters are set as  $\delta=-t$, $\omega=t$, $\lambda=0.5t$ in panel (c) and $\delta=2t$, $\omega=t$, $\lambda=0.5t$ in panel (d). The lattice size is fixed as  $L_x\times L_y=40\times 40$.}}\label{symmetry}
\end{figure}

 In the above discussion, we show that the emergence of topologically protected zero-energy corner states is intrinsically related to the NNNN anisotropic coupling. If the NNNN coupling is replaced by NN anisotropic coupling, SOTI will \rev{consequently} be replaced by other interesting topological phenomena. In this section, we will discuss two different types of  NN anisotropic coupling terms \rev{associated with two} different lattice structures. Different types of lattice structure, which depend on the existence of dislocation between the two spin components as shown in Fig.\,\ref{symmetry}, exhibit very different topological properties. In the first case, the corner states still exist but their energy is shift away from zero, and they are non-topologically protected states. In the second case, the system becomes a topological semimetal.

\subsection{Spin-independent lattice structure}
\begin{figure}[tph]
\centering
\includegraphics[width=84mm]{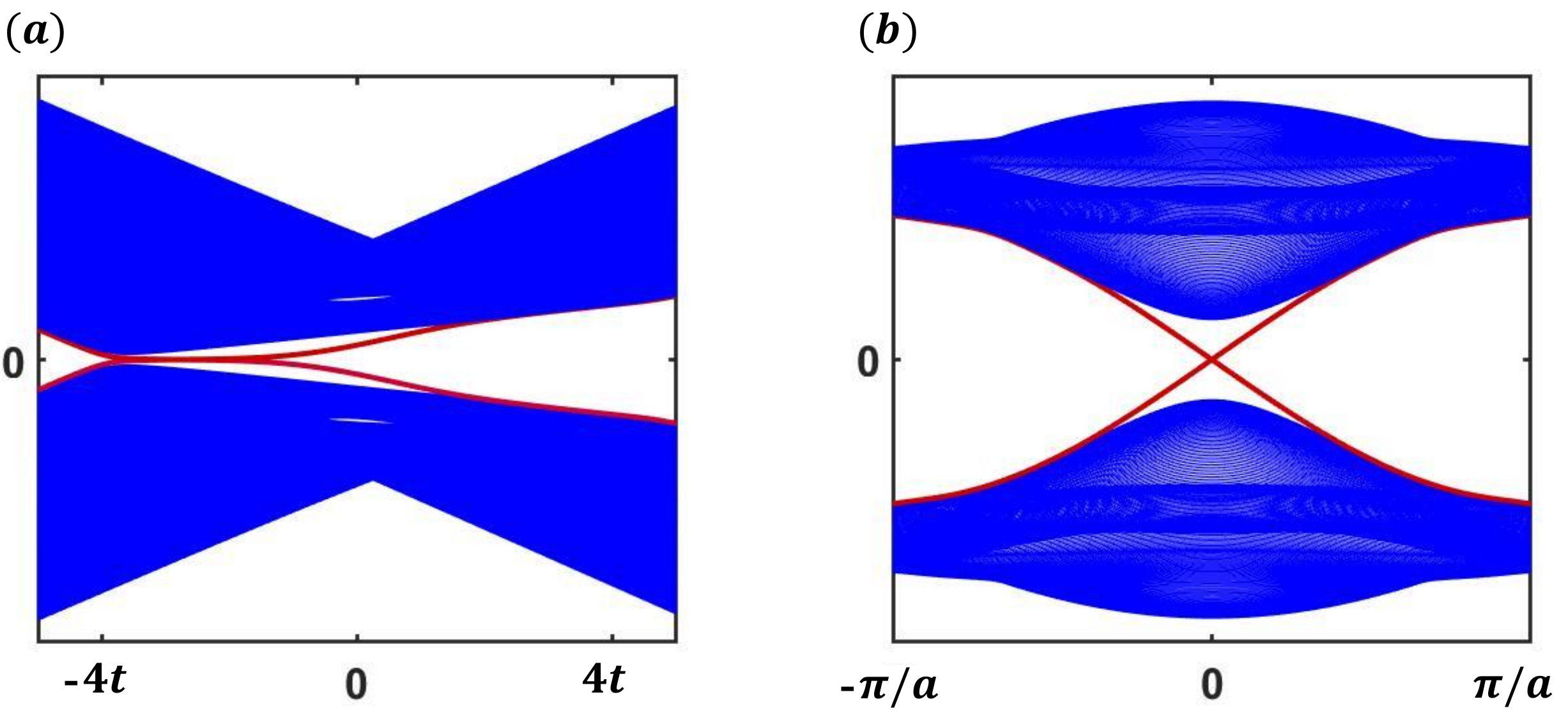}
\caption{(a) Energy spectrum \rev{of Hamiltonian (\ref{effhamwithoutglide})} as the function of the effective energy detuning $\delta$. (b) Energy-band structure of Hamiltonian (\ref{effhamwithouttimereversal}) with \rev{an} open boundary along \rev{the} $y$ direction. The edge states are gapless with NN anisotropic coupling term of the second form. \rev{The parameters are chosen as $\omega=t$, $\lambda=t$ for panel (a) and  $\omega=t$, $\lambda=0.3t$ for panel (b)}.}\label{fig:breakglide}
\end{figure}

We first consider the lattice structure without dislocation for different spin components, as shown in Fig.\,\ref{symmetry}(a). The NN coupling term is expressed as
\begin{equation}
  H_c=2\lambda\big\{\cos(k_x a)-\cos(k_ya)\big\}\sigma_x s_x
  \label{couplingtermwithoutglide}
\end{equation}
The tight-binding Hamiltonian in the momentum space is written as
\begin{equation}
\begin{split}
H^{(1)}_{\bf k}=&\big\{\delta+4t\cos(k_x a)\cos(k_ya)\big\}\sigma_z+2\Omega\sin(k_x a)\sigma_y\\
+&2\Omega\sin(k_ya)\sigma_x s_z+2\lambda(\cos(k_x a)-\cos(k_ya)\sigma_x s_x
\label{effhamwithoutglide}
\end{split}
\end{equation}
By numerical calculations with open boundary along \rev{the} $x$ and $y$ directions, we find that the energy of corner states \rev{shifts} away from zero, see Fig.\,\ref{fig:breakglide}(a). Contrary to the result of the above section, the energy band has only one crossing point at $\Gamma$ when $\delta=-4t$ and the system is gapped in the whole BZ for other values of $\delta$, \rev{implying} that the system is \rev{topologically} trivial and the corner states are non-topologically protected.

The Hamiltonian (\ref{effhamwithoutglide}) has a similar form with \rev{the} Hamiltonian (\ref{effhamsocmanybody}), \rev{except} that the NNNN coupling term is replaced by the NN coupling. But in this case, the Hamiltonian
in the folded BZ breaks the reflection symmetry, see the lattice structure in Fig.\,\ref{fig:breakglide}(a), which means that the edge polarization is not quantized.  The topological properties of this system can also be explained by \rev{the} effective edge theory.  We expand the Hamiltonian around the \rev{point with a minimum band gap}. According to the low-energy effective edge theory, we derive the effective Hamiltonian at the \rev{minimum-gap point} $\Gamma$ where $\delta\rightarrow 4t$ that is the same as the low-energy Hamiltonian in Ref.\,\cite{ZhongboYan}, ensuring the existence of corner states. But when $\delta\rightarrow -4t$, the minimum of the gap is at $M$ point and the Hamiltonian (\ref{effhamwithoutglide}) expanded around $M$ point reads
 \begin{equation}
\begin{split}
H^{(1)}_M({\bf k})=&\big\{m'+2ta^2(k^2_x+k^2_y)\big\}\sigma_z \\
-2\Omega a&\big(k_x\sigma_y-k_y\sigma_x s_z\big)-\lambda\big[4-a^2(k^2_y+k^2_x)\big]\sigma_x s_x
\label{approachformxnncoupling}
\end{split}
\end{equation}
 The effective coupling term is a constant with \rev{an} amplitude of $4\lambda$, which indicates that the system is topologically trivial. One can find that the corner states are non-topologically protected in the parameter space of $\lvert \delta \rvert<4t$, as shown in Fig.\ref{fig:breakglide}(a).

\subsection{Spin-dependent lattice structure}
\begin{figure}[tph]
\centering
\includegraphics[width=84mm]{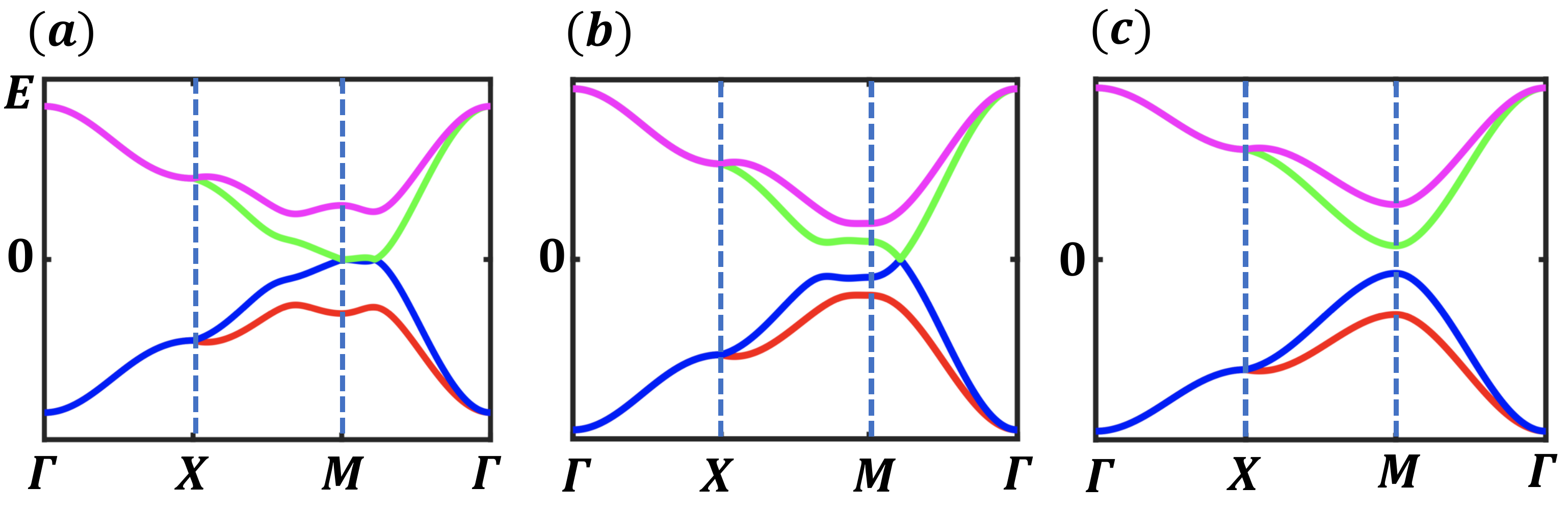}
\caption{Band structures of the Hamiltonian (\ref{effhamwithouttimereversal}) for different values of $\delta$.  (a) \rev{For $\delta=2.8t$,} the band structure has three \rev{band-crossing} points. \rev{In particular,} the band \rev{at the crossing point $M$} has \rev{a} quadratic dispersion. (b) \rev{For $\delta=3.6t$}, there is a pair of band crossing points with a linear dispersion along the $k_y=0$ line. (c) \rev{For $\delta=6.0t$,} it is a topologically trivial phase and no band touching point occurs. \rev{Here other parameters are set as $\Omega=0.8t$ and $\lambda=0.3t$}.}
\label{fig.sm}
\end{figure}

Next we consider the lattice structure with a dislocation between different spin components as shown in Fig.\ref{symmetry}(b). In this case, the NN coupling term is
 \begin{equation}
  H_c=2\lambda\big\{\cos(k_x a)-\cos(k_ya)\big\}s_x
  \label{couplingtermwithglide}
\end{equation}
The tight-binding Hamiltonian in the momentum space is
\begin{equation}
\begin{split}
H^{(2)}_{\bf k}=&\big\{\delta+4t\cos(k_x a)\cos(k_ya)\big\}\sigma_z+2\Omega\sin(k_x a)\sigma_y \\
+&2\Omega\sin(k_ya)\sigma_x s_z+2\lambda(\cos(k_x a)-\cos(k_ya)s_x
\label{effhamwithouttimereversal}
\end{split}
\end{equation}
The anisotropic term breaks the combination $\mathcal{I}\mathcal{T}$ of the inversion symmetry and the time-reversal symmetry, resulting in a splitting of the two-fold degeneracy of energy bands, as depicted in Fig.\,\ref{fig.sm}. The Hamiltonian in the folded BZ still respects the reflection symmetry, but the corresponding symmetry operators along the $x$ and $y$ directions become the form of $M_x=\sigma_z\tau_y s_y$ and $M_y=\sigma_x s_x$ respectively. They \rev{commute} with each other, which implies that there is no higher-order topological states in this case \cite{Bernevig1,Bernevig2}. By numerical calculations with \rev{a} square geometry, we find that the topologically protected zero-energy corner states disappear, see Fig.\,\ref{symmetry}(d). The band structure with \rev{an} open boundary along the $y$ direction is still gapless when $\lambda\ne 0$,  see Fig.\,\ref{fig:breakglide}(b). This can also be derived from the effective edge theory. We can expand the Hamiltonian around $\Gamma$ point when $\delta\rightarrow -4t$
\begin{equation}
\begin{split}
H^{(2)}_\Gamma({\bf k})=&\big\{m-2ta^2(k^2_x+k^2_y)\big\}\sigma_z \\
+2\Omega a&\big(k_x\sigma_y+k_y\sigma_x s_z\big)+\lambda a^2(k^2_y-k^2_x)s_x
\label{approachformgammainham2}
\end{split}
\end{equation}
\rev{When} considering the open boundary condition $y>0$, this effective Hamiltonian is separated into two parts as $H^{(2)}_\Gamma(k_x,-\partial_y)=H^{(2)}_0+H^{(2)}_p$ where $H^{(2)}_0$ is same \rev{as in} Eq.\,(\ref{edgeham0}) and $H^{(2)}_p$ can be written as
 \begin{equation}
H^{(2)}_p=2\Omega a k_x\sigma_y-\lambda a^2\partial^2_y s_x
\end{equation}
We project $H^{(2)}_p$ into the subspace consisting of the two zero-energy solutions of  Eq.(\ref{zeroenergystate0}) \rev{and find}
\begin{equation}
H^{(2)}_{y+}(k_x)\propto D_{y+}k_x s_z
\end{equation}
\rev{We remark that} the last term in $H^{(2)}_p$ can not gap out the edge state, see Fig.\ref{fig:breakglide}(b). This result \rev{leads} to the density distribution of wave functions as shown in Fig.\,\ref{symmetry}(d) for a real space lattice.

Although the Hamiltonian (\ref{effhamwithouttimereversal}) do not support HOTI and corner states, it corresponds to a topological semimetal \cite{semimetal1,semimetal2,semimetal3}. If the amplitude of $\lambda$ is large enough, there exists one or two pairs of nodal points along the line $k_y=0$ when the values of $\delta$ are in a certain range as shown in Fig.\,\ref{fig.sm}. On the $k_y=0$ line, the Hamiltonian (\ref{effhamwithouttimereversal}) \rev{can be separated} into two independent parts which include $\boldsymbol{\sigma}$ and $\boldsymbol{s}$ terms. The corresponding energy spectrum is $E(k_x)=\pm\sqrt{\big\{\delta+4t\cos(k_x d)\big\}^2+4\Omega^2\sin^2(k_x d)}\pm 2\lambda\big\{1-cos(k_x d)\big\}$. The solution to the equation $E(k_x)=0$, if existing, will correspond to the position of nodal points. Two of the phase boundaries are estimated as $\delta_c=4(t\pm \lambda)$, which implies that the band crossing point at $M$ has a quadratic dispersion as $~\frac{\Omega^2}{2\lambda}(k^2_x+k^2_y)$. \rev{For} $\delta>4(t+\lambda)$, the system is topologically trivial and has no band touching points; while for $\delta<4(t+\lambda)$, the crossing point of this band splits into two nodal points with a linear dispersion which move towards the opposite directions along  the $k_y=0$ line, as shown in Fig.\,\ref{fig.sm}(c) and (b) respectively. At $\delta=4(t-\lambda)$, another crossing point with a quadratic dispersion relation emerges at the $M$ point, see Fig\,\ref{fig.sm}(a), which similarly splits into another pair of nodal points if $\delta<4(t-\lambda)$. At another critical point, these two pairs of nodal points merge with each other and thus disappear, which further results in a topologically trivial system. We note that the value of $\lambda$ at such a critical point depends on the values of other parameters in this system.

\section{Chiral hinge modes in three-dimensional systems}
\begin{figure}[tph]
\centering
\includegraphics[width=84mm]{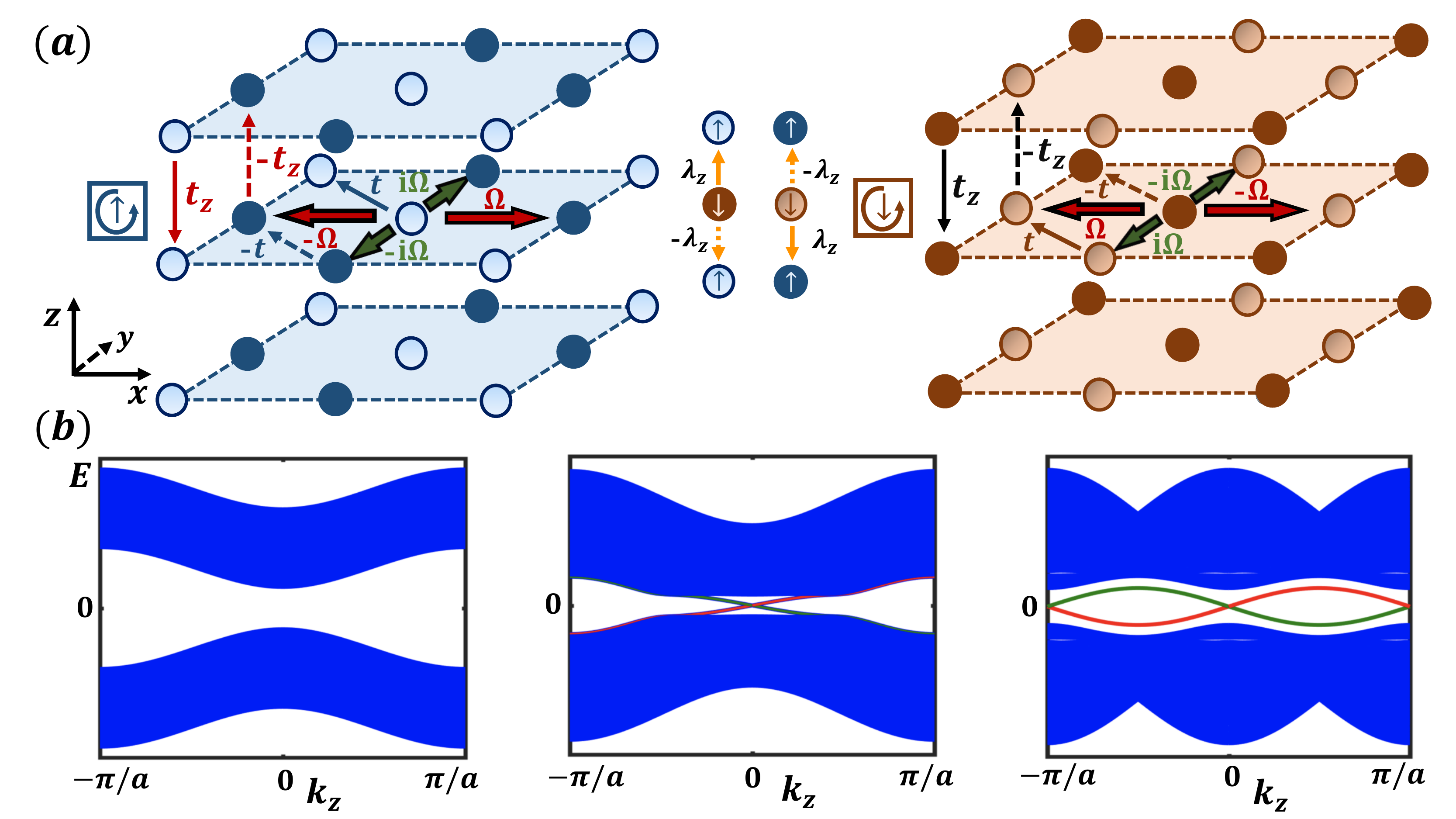}
\caption{(a) Scheme of \rev{the} 3D SOTI. (b) Chiral hinge modes in \rev{the} 3D system with open boundaries along the $x$ and $y$ directions. From the left to right panels, the energy detunings between $A$ and $B$ sublattices are $\delta=-8t$, $\delta=-4t$ and $\delta=0$ respectively, which correspond to zero, one and zero chiral hinge modes on each corner. \rev{Particularly,} the corner states swing at each corner when $\delta=0$.}
\label{hingemode}
\end{figure}

Our system can be extended to 3D system by \rev{directly} stacking 2D lattices along the $z$ direction as shown in Fig.\ref{hingemode}(a). By adding an additional NN tunneling and coupling along the $z$ direction,  one can construct a spin-dependent 3D checkerboard lattice with the Hamiltonian in the momentum space \rev{given by}
\begin{equation}
\begin{split}
H^\mathrm{3D}_{\bf k}&=\big\{\delta+4t\cos(k_x a)\cos(k_ya)+2t_z\cos(k_z a)\big\}\sigma_z \\
+&2\Omega\sin(k_x a)\sigma_y+2\Omega\sin(k_ya)\sigma_x s_z \\
+&2\lambda\big\{\cos(2k_x a)-\cos(2k_ya)\big\}\sigma_x s_x+2\lambda_z\sin(k_z a)\sigma_x s_y
\end{split}
\label{ChiralHingeMode3D}
\end{equation}
\rev{When} considering a periodic boundary condition along the $z$ direction and open boundary with a square cross section in the $x$-$y$ plane, the change of chiral hinge modes can be observed in this system, which \rev{is related with} a topological phase transition. As shown in Fig.\,\ref{hingemode}, we just consider the case \rev{of} $t_z=t$ and $\lambda_z=\lambda$ to simplify our discussion. As for $|\delta|>6t$, the system is topologically trivial and no chiral hinge modes emerge. But when $2t<|\delta|<6t$, there is one chiral mode along each hinge. When $|\delta|<2t$, the corner states swing at each corner and have no chirality.

The topological properties of this system can be \rev{viewed} as an effective 2D lattice system with an additional parameter $k_z$. In this case, the effective energy detuning of \rev{the} 2D system is $\delta_\mathrm{3D}=\delta+2t\cos(k_z a)$, which is dependent on $k_z$. \rev{For} $|\delta|>6t$, the effective energy detuning $|\delta_\mathrm{3D}|>4t$, which implies that the system is topologically trivial. \rev{While for} $|\delta|<2t$, the effective energy detuning $|\delta_\mathrm{3D}|<4t$, \rev{indicating} that the system is always topologically non-trivial. \rev{As} compared with the original 2D Hamiltonian (\ref{effhamsocmanybody}), the constant additional term $2\lambda_z\sin(k_z a)\sigma_x s_y$, which breaks the chiral symmetry, lifts zero energy of the corner states. When $2t<|\delta|<6t$, the effective energy detuning $\delta_\mathrm{3D}$ varies from the topologically trivial to the non-trivial areas with the change of $k_z$, and the chiral hinge modes emerge.

\section{The scheme of experimental setup}

{\color{black} Our scheme can be realized using periodically driven lattice. In our earlier work, we propose a scheme to realize an effective Hamiltonian \cite{slzhang}
\begin{equation}\label{shakenQAHE}
\begin{split}
 H_{{\bf k},+}=&\big\{\delta+4t\cos(k_x a)\cos(k_ya)\big\}\sigma_z \\
 +&2\Omega\sin(k_x a)\sigma_y+2\Omega\sin(k_ya)\sigma_x
\end{split}
\end{equation}
which is the left-up block of Hamiltonian (\ref{effhamsocmanybody}). \rev{We first load} ultracold atoms in a 2D checkerboard superlattice with a lattice potential $V({\bf r})=-V_0\big\{\cos^2(k_0x)+\cos^2(k_0y)+2\cos\alpha\cos(k_0x)\cos(k_0y)\big\}$, where $V_0$ and $\alpha$ are potential depth and the angle of polarization respectively, which can be well controlled in \rev{real} experiments, $k_0$ is the wave vector of the laser. The finite energy detuning prohibits the tunneling between NN sites. Then \rev{we adiabatically add a periodical} circular driving on the checkerboard superlattice with the form $x\rightarrow x+f\cos(\omega t)$, $y\rightarrow y+f\sin(\omega t)$, where $f$ and $\omega$ denote the driving amplitude and frequency respectively. In the Floquet framework, the effective Hamiltonian can be written as $H-i\hbar\partial_t=[{\bf p}-{\bf A}(t)]^2/(2m)+V({\bf r})-i\hbar\partial_t$, where ${\bf A}(t)=(f m\omega\sin(\omega t),f m\omega\cos(\omega t))$ can be considered as a \rev{temporally} periodic synthetic gauge field which can induce an effective photon-assisted NN chiral tunneling.}

\begin{figure}[tph]
\centering
\includegraphics[width=0.46 \textwidth]{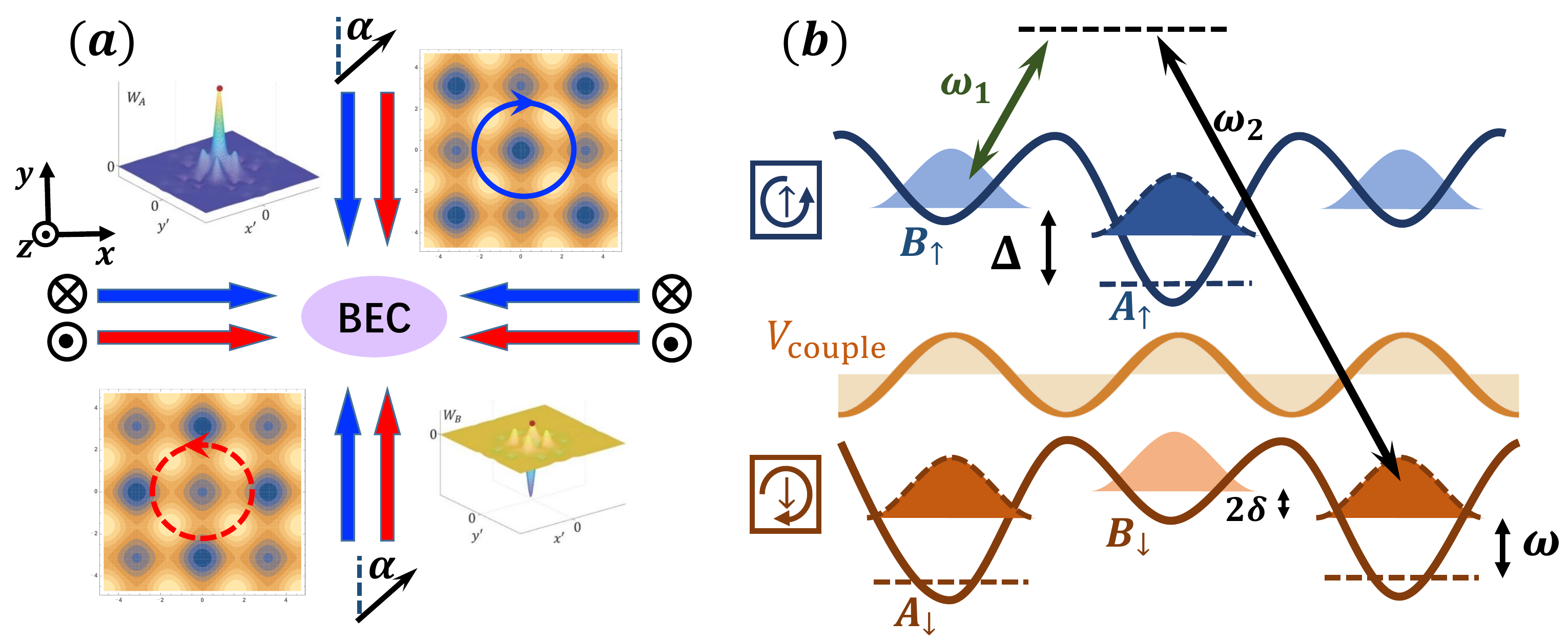}
\caption{Scheme of the experimental setup. (a) Realization of spin-dependent checkerboard lattices. Blue and red lines correspond to lasers trapping spin-up and spin-down components respectively. The polarizations of blue and red lasers \rev{along} the $x$ direction are parallel to the $z$ direction but anti-parallel to each other, while \rev{the polarizations of lasers along} the $y$ direction are parallel to each other but have an angle $\alpha$ with the $z$ axis. The right upper and left lower insets are the lattice structures for spin-up and spin-down components. The clockwise and counter-clockwise circle correspond to the chirality of periodic driving.\rev{The left upper and right lower insets depict} the Wannier functions of $A$ and $B$ sublattices respectively  (b) The brown pattern is the form of coupling between different spin components. Red and purple wave packets correspond to the Wannier wave functions for the spin-up and spin-down components respectively. {\color{black} where $x'$ and $y'$ correspond to $x+y$ and $x-y$ directions. The red dots show the positions of $(x,y)=(0,0)$, which means that the centers of two different types of Wannier functions have a displacement of $a$.} }
\label{experimentalsetup}
\end{figure}

{\color{black} To realize the Hamiltonian (\ref{effhamsocmanybody}), we need to \rev{exploit two hyperfine spin states} of ultracold atoms. Using two types of lasers with different frequencies, \rev{atoms of the different two hyperfine states can be trapped separately \cite{spindependent1,spindependent2}.} By controlling the angle of polarization for different lasers, the spin-dependent checkerboard lattice can be realized, as shown in Fig.\,\ref{experimentalsetup}(a). We assume \rev{that the frequency difference of lasers} does not induce visible difference of the wavelengths, and the lattice spacing for different spin components are approximately the same. Then by adiabatically adding periodically circular driving with opposite rotation directions for different hyperfine spin components, the effective Hamiltonian (\ref{effhamsocmanybody}) with $\lambda=0$ can be simulated.

To simulate the off-diagonal terms which \rev{depend} on $\lambda$, one can try to \rev{utilize} an additional magnetic field along the $z$ direction and a two-photon Raman process. \rev{Such a} magnetic field induce an energy split $\Gamma$ between different spin components. Two Raman lasers with \rev{the} same wave vector $k_0$ and an frequency difference $\omega_2-\omega_1=\Gamma$ can be set as in Fig.\,\ref{experimentalsetup}(b), which \rev{implies that} the Raman coupling is inhomogeneous with the form $V_\mathrm{couple}(x,y)=\Lambda \big\{\cos(2k_0y)-\cos(2k_0x)\big\}|\uparrow\rangle\langle\downarrow|$.\cite{zwu,zywang} The minus sign can be controlled by setting the relative phase between \rev{the} two Raman lasers. \rev{Based on} this setting, the on-site coupling will be cancelled because the Wannier function is symmetric along the $x$ and $y$ direction. The NN and NNNN couplings along the $x$ direction can be estimated as
\begin{equation}
\begin{split}
\lambda_{NN}=&\int W^*_A(x,y)V_\mathrm{couple}(x,y)W_B(x,y)dx dy \\
\lambda_{NNN}=&\int W^*_A(x,y)V_\mathrm{couple}(x,y)W_B(x+a,y+a)dx dy \\
\lambda_{NNNN}=&\int W^*_A(x,y)V_\mathrm{couple}(x,y)W_A(x+2a,y)dx dy
\end{split}
\end{equation}
where $W_A(x,y)$ and $W_B(x,y)$ are the Wannier functions of $A$ and $B$ sublattices of the original checkerboard lattices respectively, as shown in {\color{black}the inset of} Fig.\,\ref{experimentalsetup}(a), the centers of two types of Wannier function have a displacement of $a$. We can verify that the amplitude of NNN coupling $\lambda_{NNN}$ along the $x\pm y$ direction can be ignored because of the orthogonality of Wannier functions. \rev{According to} the form of Raman coupling, the NN and NNNN couplings along the $y$ direction have the same amplitude but opposite signs \rev{as compared to the ones along the} $x$ direction. From the form of Wannier functions, the amplitude of $\lambda_{NN}$ always has the same order \rev{as} $\lambda_{NNNN}$. To overcome the influence of $\lambda_{NN}$, we can engineer the form of Raman coupling more delicately. But actually we don't need to worry about the influence of $\lambda_{NN}$. From our discussion of the second case in section III, this type of NN coupling can not open the gap on the boundary, which means that it has no contribution to the higher-order topology. {\color{black} This result can also be verified by numerical calculation.}
 Using the experimental feasible lattice potential $V_0=2E_R$($E_R=h^2/(8m a^2)$ is the recoil energy), $\alpha=0.02$ and \rev{suitable} Raman coupling strength, one can estimate that the amplitude of $\lambda_{NNNN}$ is about $0.01E_R$ when $\Lambda$ is on the same order of $E_R$, which is on the order of $nK$. \rev{According to} our calculation, the energy gap always depends on the amplitude of effective chiral tunneling induced by the lattice shaking when the amplitude of $\lambda_{NNNN}$ is small. So we think all phenomena we discussed in our work can be observed in current experiments.

\rev{One} can \rev{make use of several} different methods to observe the topological properties in this system. The most direct method is \rev{to create} a sharp open boundary on the $x$ and $y$ directions and measure the corner states directly by using the optical box as in Ref.\cite{zoran}{\color{black} to create a sharp boundary}. But in ultracold atom systems, we can also try to detect the topological properties \rev{of} the bulk: (i) The first method is using the topological pumping. We can \rev{regard $k_z$ in Hamiltonian (\ref{ChiralHingeMode3D}) as a time-dependent parameter} and try to change \rev{it} adiabatically. Then the movement of particles towards two diagonal corners can be observed \cite{cornerpump}; (ii) The second method is the direct measurement of the Wilson loop. By accelerating the optical lattice and {\color{black} with} band mapping, one can realize the tomography of Berry curvature and Wilson loop \cite{tomo1,tomo2}. Then we can reconstruct the quantized quadrupole moment in the bulk.}

\section{Conclusion}

As a conclusion, we investigate a second-order topological insulator with open boundary along special direction. With the NNNN anisotropic coupling, zero-energy corner states emerge. {\color{black} We} find that nested wilson loop can also be well defined {\color{black} in the system with reflection symmetry even when the Wannier bands are gapless}. We also extend our system to \rev{a} 3D system and propose a 3D second-order topological insulator with chiral hinge modes. We hope our work will stimulate more explorations about the choice of open boundaries and the properties of higher-order topological quantum matters. In future, we will try to search for higher-order topological insulator systems which can be simulated more easily in real experiments. We will also extend our system into higher dimensions and explore more interesting quantum phenomena because of the interplay between symmetries and high-order topological properties.

\section*{ACKNOWLEDGEMENTS}

We thanks Xiang-Fa Zhou for his valuable suggestions. S. L. Zhang is supported by the Program of State Key Laboratory of Quantum Optics and Quantum Optics Devices (No:KF201903).

\end{document}